\documentclass{appolb}
\usepackage{epsfig}
\usepackage{bbm}
\newcommand{\lesssim}{\:\mbox{\raisebox{-3pt}{$\stackrel%
{\displaystyle <}{\sim}$}}\:}
\newcommand{\gtrsim}{\:\mbox{\raisebox{-3pt}{$\stackrel%
{\displaystyle >}{\sim}$}}\:}
\newcommand{\datm}{\Delta m^2_\mathrm{atm}}
\newcommand{\dsol}{\Delta m^2_\mathrm{sol}}
\begin{document}
\title{Renormalisable $SO(10)$ models and \\
neutrino masses and mixing
\thanks{Presented by W.~Grimus at XXXI International Conference of Theoretical
  Physics, Matter to the Deepest: Recent Developments in Physics of
  Fundamental Interactions, Ustro\'n, 5--11 September 2007, Poland}
}
\author{Walter Grimus
\and
Helmut K\"uhb\"ock
\address{University of Vienna, Faculty of Physics \\
Boltzmanngasse 5, A-1090 Vienna, Austria}
}
\maketitle
\begin{abstract}
We discuss some recent developments in SUSY Grand Unified Theories
based on the gauge group $SO(10)$. Considering renormalisable Yukawa
couplings, we present ways to accommodate quark and lepton masses and 
and mixings.
\end{abstract}
  
\section{Introduction}
Grand unified theories (GUTs) based on 
the group $SO(10)$~\cite{fritzsch} are interesting because 
its 16-dimensional irreducible
representation (irrep), the spinor representation, contains all chiral
fermions included in a Standard Model (SM) family plus an additional 
SM gauge singlet, the right-handed neutrino. Moreover, such theories allow 
type~I and type~II seesaw mechanisms for generating light neutrino masses. 
The basis of the Lie algebra $so(10)$ consists of 45 antisymmetric real
matrices, usually taken to be
\begin{equation}
(M_{pq})_{jk} = \delta_{pj} \delta_{qk} - \delta_{qj} \delta_{pk}
\quad (1 \leq p < q \leq 10),
\end{equation}
i.e.,
\begin{equation}
M_{12} = \left( \begin{array}{ccc}
0 & 1 & \cdots \\
-1 & 0 & \cdots \\
\vdots & \vdots & \ddots
\end{array} \right), \quad \mbox{etc.}
\end{equation}
The commutation relations are
\begin{equation}
[M_{pq},M_{rs}] = \delta_{ps} M_{qr} + \delta_{qr} M_{ps} -
\delta_{pr} M_{qs} - \delta_{qs} M_{pr}.
\end{equation}
Defining a Clifford algebra with basis elements $\Gamma_p$ and
anticommutation relations
\begin{equation} 
\{ \Gamma_p,\Gamma_q\} = 2 \delta_{pq} \mathbbm{1}  \quad
(1 \leq p,\,q \leq 10),
\end{equation}
it is easy to show that the quantities~\cite{sakita,wilczek}
\begin{equation}\label{sigma}
{\textstyle \frac{1}{2}} \sigma_{pq} \equiv 
{\textstyle \frac{1}{4}} [\Gamma_p,\Gamma_q]
\end{equation}
fulfill the $so(10)$ commutation relations. Therefore, any 
representation of the Clifford algebra
is at the same time a representation of $so(10)$. 

\subsection{The 16-dimensional spinor representation}

The Clifford algebra associated with $so(10)$ has a 32-dimensional
irrep. However, the associated $so(10)$ representation
obtained via~(\ref{sigma}) is reducible. It decays into the
spinor irrep $\mathbf{16}$ and its complex conjugate 
$\overline{\mathbf{16}}$. The Pati-Salam group 
$G_{422} \equiv SU(4)_C \times SU(2)_L \times SU(2)_R$, where
$SU(4)_C$ unifies colour and lepton number~\cite{PS}, is very useful
for a classification of the fields contained in $so(10)$ irreps. The
decomposition of the $\mathbf{16}$ is given by
\begin{equation}
\mathbf{16} \stackrel{422}{=} 
(\mathbf{4},\mathbf{2},\mathbf{1})\oplus
(\overline{\mathbf{4}},\mathbf{1},\mathbf{2}).
\end{equation}
The further decomposition of the $G_{422}$ multiplets with respect to the 
SM gauge group $G_{321} \equiv SU(3)_c \times SU(2)_L \times U(1)_Y$
is
\begin{eqnarray}
(\mathbf{4},\mathbf{2},\mathbf{1})
& \stackrel{321}{=} &
(\mathbf{3},%
\mathbf{2})_{1/6} \oplus (\mathbf{1},\mathbf{2})_{-1/2}\,, \\
(\overline{\mathbf{4}},\mathbf{1},\mathbf{2})
& \stackrel{321}{=} & 
(\overline{\mathbf{3}},\mathbf{1})_{1/3} \oplus
(\overline{\mathbf{3}},\mathbf{%
1})_{-2/3} \oplus (\mathbf{1},\mathbf{1})_{1} \oplus
(\mathbf{1},\mathbf{1})_{0}\,.
\end{eqnarray}
Thus the SM fermion field assignments (all fields are to be considered
left-handed) are given by 
\begin{equation}
(\mathbf{4},\mathbf{2},\mathbf{1}):\left(
\begin{array}{llll}
u_{r} & u_{y} & u_{b} & \nu \\
d_{r} & d_{y} & d_{b} & e %
\end{array}%
\right) ,\,\,(\overline{\mathbf{4}},\mathbf{1},\mathbf{2}):\left(
\begin{array}{llll}
d_{r}^{c} & d_{y}^{c} & d_{b}^{c} & e^{c} \\
u_{r}^{c} & u_{y}^{c} & u_{b}^{c} & \nu^{c}%
\end{array}%
\right).
\end{equation}

\subsection{Scalars for Yukawa Couplings}
For the Yukawa couplings one has two options: One option is to take into
account only ``low-dimensional'' scalar irreps like $\mathbf{10}$ and 
$\mathbf{16}$; in that case one has to resort to non-renormalizable
interactions. The other option~\cite{sakita,wilczek} 
is to take the scalar irreps which appear in
\begin{equation}\label{16x16}
\mathbf{16} \otimes \mathbf{16} =
{10 \choose 1} \oplus {10 \choose 3} \oplus \frac{1}{2}%
{10 \choose 5} = \mathbf{10} \oplus \mathbf{120} \oplus \mathbf{126}\,;
\end{equation}
here one has rather high-dimensional irreps. Note that in 
$\mathbf{16} \otimes \mathbf{16}$ 
only totally antisymmetric tensors with uneven rank occur:
the vector $\mathbf{10}$, 
the totally antisymmetric 3-tensor $\mathbf{120}$ and 
the totally antisymmetric, selfdual 5-tensor $\mathbf{126}$. Selfdual
means that $\ast\mathbf{126}=i\mathbf{126}$; the star indicates the
formation of the dual tensor with the epsilon tensor of rank 10.
The $\mathbf{126}$ is a genuinely complex irrep. Therefore, the 
scalars for renormalisable Yukawa couplings are given by
$\mathbf{10}$, $\mathbf{120}$ and $\mathbf{\overline{126}}$. 

\subsection{The Yukawa Lagrangian and fermion mass matrices}

With the scalar irreps of the previous subsection the Yukawa
Lagrangian reads~\cite{sakita}
\begin{eqnarray}
\mathcal{L}_{Y} &=&
\frac{1}{2} \left( H_{ab} \mathbf{16}_{aL}^{T}C^{-1}\mathcal{B}\,%
\Gamma_{p}\mathbf{10}_{p}\mathbf{16}_{bL}+ 
\right. \\ && \hphantom{xxi}
G_{ab} \mathbf{16}_{aL}^{T} C^{-1} 
\mathcal{B}\, \Gamma_{p}\Gamma_{q}\Gamma_{r}
\mathbf{120}_{pqr} \mathbf{16}_{bL} + \\
&& \left. \hphantom{xxi}
F_{ab} \mathbf{16}_{aL}^{T}C^{-1} \mathcal{B}\, 
\Gamma _{p}\Gamma_{q}\Gamma _{r}\Gamma _{s}\Gamma
_{t}\overline{\mathbf{126}}_{pqrst}\mathbf{%
16}_{bL} + \mbox{H.c.} \right).
\end{eqnarray}
In this Lagrangian we have
$SO(10)$ indices $1 \leq p,q,r,s,t \leq 10$ and 
family indices $1 \leq a,b \leq 3$. The 
$SO(10)$ ``charge-conjugation matrix'' $\mathcal{B}$ has the properties
\begin{equation}
\mathcal{B}^T = \mathcal{B}, \quad
\mathcal{B}^{-1}\Gamma_{p}^{T}\mathcal{B}=\Gamma_{p}.
\end{equation}
Due to the structure of $\mathcal{L}_Y$, the Yukawa coupling matrices
must fulfill
\begin{equation}
H^{T}= H,\quad G^{T}= -G,\quad F^{T}= F.
\end{equation}
The vacuum expectation values (VEVs) 
$k_{d,u}$, $\kappa_{d,u,\ell,D}$, $v_{d,u}$ 
of the Higgs doublets contained
in the scalars of $\mathcal{L}_Y$ determine the mass matrices of the 
fermions. For the charged fermions we have 
\begin{eqnarray}
M_{d} & = & k_{d}\,H+\kappa_{d}\,G+v_{d}\,F, \label{Md} \\
M_{u} & = & k_{u}\,H+\kappa_{u}\,G+v_{u}\,F, \label{Mu} \\
M_{\ell} & = & k_{d}\,H + \kappa_\ell \,G-3\,v_{d}\,F. \label{Ml} 
\end{eqnarray}
The $-3$ is a Clebsch--Gordan coefficient. In the neutrino sector also
$SU(2)$ triplet VEVs $w_R$ and $w_L$, stemming from 
the $\overline{\mathbf{126}}$, occur. One needs the following matrices:
\begin{equation}
M_{D} = k_{u}\,H + \kappa_{D}\,G -3\,v_{u}\,F, \quad
M_{R} = w_{R}\,F, \quad M_{L} = w_{L}\,F,
\end{equation}
where $M_R$ with the large VEV $w_R$ is the mass matrix of the heavy
Majorana neutrinos. The mass matrix of the light neutrinos is
determined by the seesaw mechanism:
\begin{equation}\label{Mnu}
\mathcal{M}_\nu = M_{L}-M_{D}^{T}M_{R}^{-1}M_{D}.
\end{equation}
The VEV $w_L$ is small according to the type~II seesaw mechanism.

\section{The Minimal SUSY $SO(10)$ GUT}

The Minimal SUSY $SO(10)$ GUT~\cite{MSGUT} (MSGUT) is characterized by
the following multiplets. Each fermion family resides in a 
$\mathbf{16}$, the gauge bosons are, of course, 
in the adjoint representation $\mathbf{45}$ and the scalars are given by
$\mathbf{10} \oplus \overline{\mathbf{126}} 
\oplus \mathbf{126} \oplus \mathbf{210}$. The $\mathbf{210}$ is the
totally antisymmetric tensor of rank 4.
The tasks of the different scalar multiplets are the following:
\begin{itemize}
\item
$\mathbf{10} \oplus \overline{\mathbf{126}}$ $\to$ Yukawa couplings,
\item
$\overline{\mathbf{126}} \oplus \mathbf{126} \oplus 
\mathbf{210}$ $\to$ breaking $SO(10)$ down to $G_{321}$,
\item
$\mathbf{126}$ $\to$ avoiding SUSY breaking at high scales.
\end{itemize}
The idea is that SUSY breaking is accomplished by soft terms at
$G_{321}$ stage. A further condition is that of 
\emph{minimal finetuning}~\cite{bajc}: At the electroweak scale there
are only two light Higgs doublets $H_d$, $H_u$, just the ones which
appear in the Minimal Supersymmetric Standard Model (MSSM). 
This is a non-trivial condition because each of the scalar multiplets
contains two SM doublets,
one for each hypercharge $\pm 1/2$.

However, the minimal SUSY $SO(10)$ is GUT ruled
out~\cite{garg,schwetz}. It is amazing that this theory is so
constrained that one can falsify it. In essence the reason for 
the failure of the MSGUT can be formulated in the following way:
\begin{quote}
The MSSM gauge coupling unification occurs at the scale 
$M_\mathrm{GUT}$ \emph{without} intermediate scale,
but neutrino masses via the seesaw mechanism require a scale 
$M_\mathrm{seesaw} < M_\mathrm{GUT}$.
\end{quote}
Identifying the electroweak scale with the SM VEV $v \simeq 174$ GeV
and using $\datm \sim 2.5 \times 10^{-3}$ eV$^2$ for the atmospheric
neutrino mass-squared difference, we find
\begin{equation}
v^2/M_\mathrm{seesaw} \gtrsim \sqrt{\datm}
\; \Rightarrow \; M_\mathrm{seesaw} \lesssim  6 \times 10^{14}\;
\mbox{GeV}.
\end{equation}
Thus, $M_\mathrm{seesaw}$ 
is more than one order of magnitude below 
$M_\mathrm{GUT} \simeq 2 \times 10^{16}$ GeV from the MSSM
gauge coupling unification. This is the source
of the problem.

Let us present some details. In the MSGUT, 
$G = 0$ in the mass formulas
(\ref{Md})--(\ref{Mnu}). Suppose the VEVs $k_{d,u}$, $v_{d,u}$,
$w_{R,L}$ are free parameters, then one obtains an excellent fit to
known fermion masses and mixings~\cite{schwetz}. 
The number of independent parameters in the system of mass
matrices is 21, 13 absolute values and 8 phases, whereas 
the number of observables is 18: nine
charged-fermion masses, two neutrino mass-squared differences,  
six mixing angles and one CKM phase. 
This fit is done by minimizing~\cite{schwetz} 
\begin{equation}\label{chi2}
\chi^{2}(p)=\sum_{i=1}^{n}
\left(\frac{f_{i}(p)-\bar{O}_{i}}{\sigma_{i}}\right)^{2},
\end{equation}
where $p = \{ p_1, \ldots, p_r \}$ is the set of parameters 
(in the MSGUT $r=21$) and 
the functions  $f_{i}(p)$ are the predictions for the
observables ($n = 18$ in our case). In~(\ref{chi2})
the input values $\bar{O}_{i}\pm\sigma_{i}$ have to be taken at the
GUT scale, extrapolated via the
renormalisation group equations of the MSSM from the values of the 
observables at the electroweak scale scale~\cite{das}.  
A suitable numerical procedure for finding the minimum of $\chi^2$ is the
downhill simplex method~\cite{schwetz}. 

Though the fit to fermion masses and mixings turns out to be excellent
if the VEVs are considered as free parameters, this is not the whole
story. In the MSGUT the number of terms in the scalar potential is limited
due to SUSY and the $SO(10)$ multiplet content. Therefore, 
the VEV ratios are not free but functions of 
$\tan \beta = \langle H^0_u \rangle/\langle H^0_d \rangle$ 
and the parameters of the scalar potential. Thus the fit 
in the fermionic sector constrains the scalar potential,
resulting in a light scalar $G_{321}$ multiplet 
$(\mathbf{8}, \mathbf{3})_0$ which destroys 
gauge coupling unification~\cite{garg,schwetz}.

\section{The MSGUT plus the scalar 120-plet}

The addition of the scalar $\mathbf{120}$ can release the
strain from the mass and mixing fit, allowing a 
``small'' Yukawa coupling matrix $F$ of the $\overline{\mathbf{126}}$ in order 
to enhance the value of the neutrino masses via the type~I seesaw
mechanism~\cite{aulakh} with the inverse of $F$.
In references~\cite{GK1,GK2} we have put forward the idea to add the 
$\mathbf{120}$ and to make the indentification
$w_R \equiv M_\mathrm{GUT}$,
in order to avoid light scalar multiplets except the MSSM Higgs
doublets. This identification was done by hand, therefore, fits to
fermion masses and mixings by leaving the other VEVs as free
parameters are only ``generic'' fits, without taking into account the
full theory. With the 120-plet 
no new heavy VEVs appear compared to the MSGUT; 
the $\mathbf{120}$ has \emph{two} SM Higgs doublets for each
hypercharge $\pm 1/2$ and now each 
MSSM Higgs doublet
$H_{d,u}$ is a linear combination of \emph{six} doublets.

The aim of~\cite{GK1,GK2} was to show that the idea presented above
works for fermion masses and mixings. 
We used the method of~\cite{schwetz} for the numerics.
However, due to the addition of
the $\mathbf{120}$ there is the numerical 
problem that the number of parameters is large and the downhill simplex method
becomes time consuming and less trustworthy. Therefore, it is necessary to
reduce the number of parameters, motivated by physical reasons. 
First of all, type~I seesaw will be dominant and one can neglect type~II
contributions to $\mathcal{M}_\nu$. 
Furthermore, one can assume 
real Yukawa couplings, motivated by spontaneous CP violation.
In~\cite{GK1} we reduced the number of parameters in addition by postulating 
a $\mathbbm{Z}_2$ symmetry: 
$\mathbf{16}_2 \to -\mathbf{16}_2$, 
$\mathbf{120} \to -\mathbf{120}$; with complex
VEVs there are 21 parameters, including 6 phases (Case A). In~\cite{GK2},
dispensing with a family symmetry, we assumed the VEVs of $\mathbf{10}$ and 
$\overline{\mathbf{126}}$ to be real real, but the VEVs of the $\mathbf{120}$
to be imaginary; this leads to 18 parameters (Case B).
\begin{figure}
\begin{center}
\epsfig{file=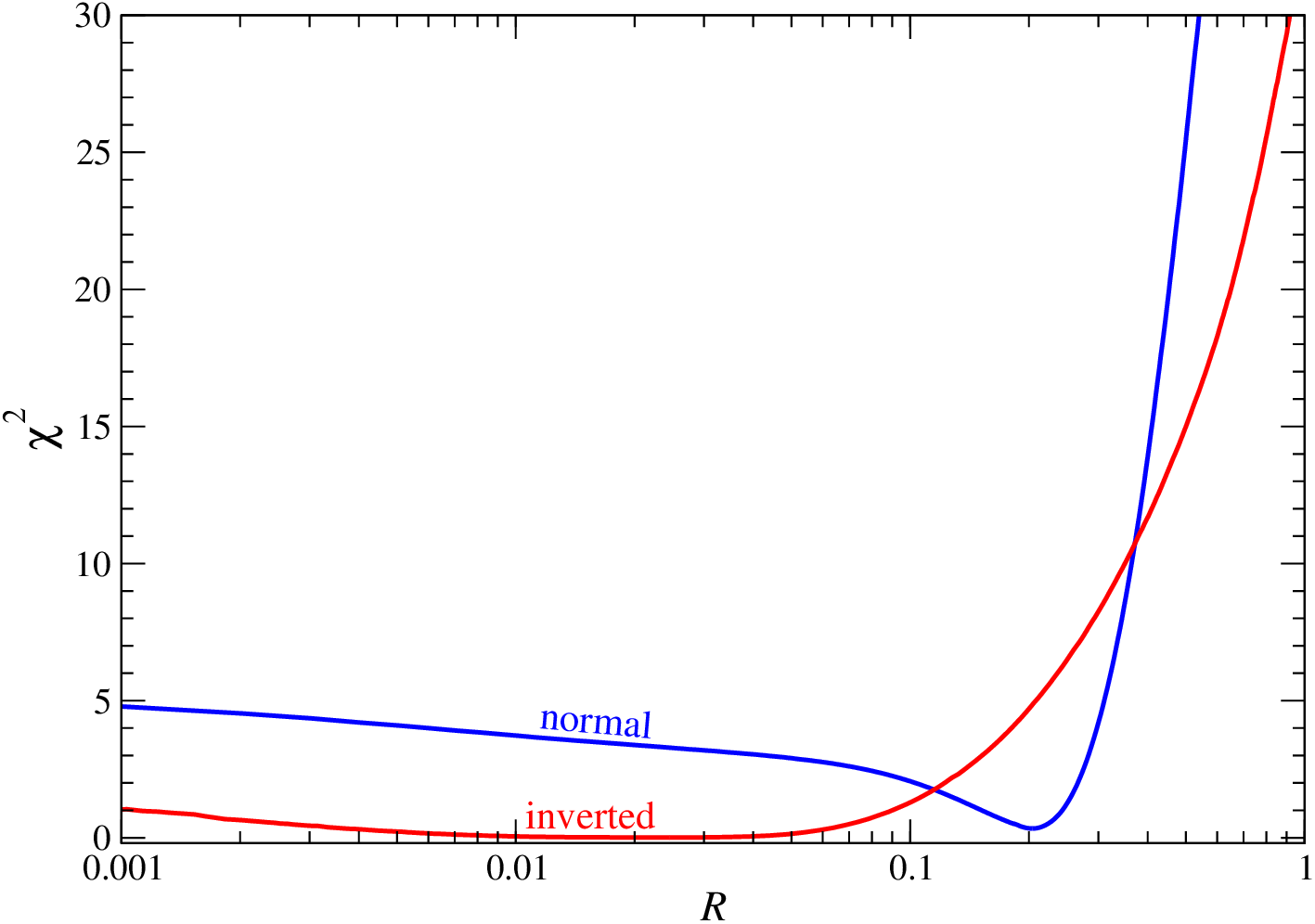,width=0.8\textwidth}
\end{center}
\caption{The $\chi^2$ of Case B as a function of 
$R = m_\mathrm{min}/\dsol$. \label{fig}} 
\end{figure}
We obtained excellent fits in both cases. 
E.g. for Case B the minimal $\chi^2$ is 0.33 with normal 
and 0.011 with inverted neutrino mass spectrum.
Unfortunately we could not find striking predictions, except that both cases 
strongly prefer a hierarchical spectrum---see Fig.~\ref{fig} where $\chi^2$ is
plotted as a function the minimal neutrino mass over the solar mass-squared
difference.

\section{Conclusions}

The MSGUT together with the scalar $\mathbf{120}$ can excellently
reproduce known fermion masses and mixings, 
while identifying the triplet VEV $w_R$ (the ``seesaw scale'') with the GUT
scale---at least by doing a generic fit without taking possible
relations among the VEVs into consideration. A more complete
discussion has been started in~\cite{garg-CP}, dubbing the model 
``New MSGUT'' or NMSGUT. Hopefully in the NMSGUT the VEVs necessary
for the fit to fermion masses and mixings are compatible with the
scalar potential on the one hand and gauge coupling unification on the
other hand. This consistency has yet to be checked.
While in the fit to the fermionic sector 
spontaneous CP violation mainly played a role in reducing the number
of parameters in the Yukawa couplings~\cite{GK1,GK2}, there are
indications that it could play a much more important role in $SO(10)$
GUTs~\cite{garg-CP,achiman}. In particular, in the NMSGUT the
requirement of spontaneous CP violation pushes the GUT scale closer to
the Planck scale~\cite{garg-CP}. In conclusion, we want to note,
however, that still there is no idea how to obtain
suitable simple fermion mass matrices 
with \emph{explanatory power} in $SO(10)$ GUTs; the
discussion still centers on accommodation of fermion masses and mixings
and consistency of the theory.

\end{document}